# Agglomeration based influential node ranking in path-type networks

[a]Zeynep Nihan BERBERLER, [b]Aysun Asena KUNT
[a,b]Faculty of Science, Department of Computer Science, Dokuz Eylul University,
35160, Izmir/TURKEY
[a]zeynep.berberler@deu.edu.tr, [b]aysunasena.kunt@ogr.deu.edu.tr

**Abstract.** Identification of vital nodes contributes to the research of network robustness and vulnerability. The most influential nodes are effective in maximizing the speed and accelerating the information propagation in complex networks. Identifying and ranking the most influential nodes in complex networks has not only theoretical but also practical significance in network analysis since these nodes have a critical influence on the structure and function of complex networks. This paper is devoted to the evaluating the importance of nodes and ranking influential nodes in paths and path-type networks such as comets, double comets, and lollipop networks by network agglomeration based node contraction method.

## 1. Introduction

Throughout this paper, we consider simple finite undirected graphs without loops and multiple links. The *order* of $G$ is the number of nodes in $G$. The *open neighborhood of* $v$ is $N_G(v) = \{u \in V(G) | uv \in E(G)\}$. The *degree* of a node $v$ in $G$ is $\deg_G(v) = |N_G(v)|$. The *maximum degree* and *minimum degree* among the nodes of $G$ is denoted by $\Delta(G)$ and $\delta(G)$, respectively. An *end-node* or a *pendant* or a *pendent node* is a node of degree one. The *distance* $d_G(u,v)$ between two nodes $u$ and $v$ in $G$ is the length of a shortest path between them. For $u = v$, $d_G(u,v) = 0$. For $S \subseteq V(G)$, the subgraph of $G$ induced by $S$ is denoted by $G[S]$. If the graph $G$ is clear from the context, we simply write $N(v)$, $\deg(v)$, and $d(u,v)$ rather than $N_G(v)$, $\deg_G(v)$, and $d_G(u,v)$, respectively [2].

Complex networks are the topological connections for systems. Complex networks are comprised of large amounts of nodes and links. In order to reveal the complexity of complex systems, the study on the relationship between the topological structures and network functions is the foundation. Hence, identifying and ranking influential nodes is a critical research task in the network science. The mining of influential nodes has both theoretical significance and practical application in complex networks such as Internet, social networks, power grids, transportation networks, chemical and biological networks, neural networks, etc. Identifying the influential nodes and estimating the spreading influence in networks are fundamental task in order to control the dissemination process [1,3,6,8-9]. Influential nodes lead to faster and wider spreading in complex networks.

A novel method of evaluating node importance by node contraction based on network agglomeration was proposed in [10] which focuses on the influence of every node by considering their contribution to the whole network. This method of identifying and ranking the most influential nodes in a complex network describes the key nodes of the network more accurately by taking into consideration not only the degree but also the position of the nodes in the network. If a node in the network is an important node which contributes to network connectivity, then we consider that the more agglomerate is the network after a node is contracted, the more important is the node.

For $v_i \in V(G)$, the node contraction is defined in [10] as node $v_i$ and other $\deg(v_i)$ nodes connected with $v_i$ into a new node $v_i'$, which takes place of the primary $\deg(v_i)+1$ nodes, and links connected with $\deg(v_i)-1$ nodes originally turn to the new node $v_i'$ now. The network after the operation of node contraction is expressed as $G'$. The high degree of agglomerate lies on two aspects: the first one is the network connectivity characterized by average path length $L$ and the second one is the network size characterized by the number of nodes $n$. The network agglomeration denoted by $\phi(G)$ is defined as the reciprocal of the product of $L(G)$ and $n$ as follows [10]:

$$\phi(G) = \frac{1}{nL(G)} = \frac{1}{n \frac{\sum_{v_i,v_j \in V(G)} d_{v_i v_j}}{n(n-1)}} = \frac{n-1}{\sum_{v_i,v_j \in V(G)} d_{v_i v_j}},$$

where $n \geq 2$ and $d_{v_i v_j}$ is the shortest path length between the nodes $v_i$ and $v_j$ $(i \neq j)$. When $n=1$, the agglomeration of the network reaches the maximum value 1, and obviously

$0 < \phi(G) \leq 1$. According to the formula of $\phi(G)$, the importance of the node $v_i$ denoted by $IMC(v_i)$ is defined as follows [10]:

$$IMC(v_i) = 1 - \frac{\phi(G)}{\phi(G'(v_i))},$$

where $G'(v_i)$ denotes the new network after contracting the node $v_i$.

By the definition of node importance, the importance of a node depends on two factors-the degree of the node and the position of the node. If the degree of a node is higher, the number of nodes after contraction is fewer, and then the network agglomeration is higher, thus the related node is more important. In other words, we can say that if a node is located in a vital position in the networks, then many shortest paths of node pairs pass through this node, thus the higher network agglomeration is obtained after this node is contracted in the network.

The method of node importance evaluation by node contraction based on network agglomeration was proved to be valid and accurate and can be applied to the ranking of influential nodes in large scale networks [10]. In addition, it is pointed out that since the contraction of node reduces the number of nodes and links, the workload of calculation decreases accordingly.

The following section is devoted to the evaluation of agglomeration of paths and path-type networks such as comets, double comets, and lollipop networks and computation of importance value of each node in the related network types. The importance analysis and ranking of each node in the related networks are conducted and analyzed in detail.

## 2. Node ranking in paths and related networks

*2.1. Path networks*

**Lemma 2.1.1.** Let $G = P_n \ (n > 1)$ be a path of order $n$. Then, the agglomeration of the path is $\phi(P_n) = 3/(n(n+1))$.

**Proof.** For the nodes of $P_n$,

$$\sum_{\forall u,v \in V(P_n), u \neq v} d_{uv} = \sum_{j=0}^{n-1}\left(\sum_{i=1}^{j} i + \sum_{i=1}^{n-1-j} i\right) = \sum_{j=0}^{n-1}\left(\frac{j(j+1)}{2} + \frac{(n-1-j)(n-j)}{2}\right)$$

$$= \frac{n(n^2-n)}{2} + \sum_{j=0}^{n-1}(j^2+j-nj) = \frac{n(n^2-n)}{2} + \sum_{j=0}^{n-1} j^2 + \sum_{j=0}^{n-1} j(1-n)$$

$$= \frac{n(n^2-n)}{2} + \frac{n(n-1)(2n-1)}{6} + (1-n)\left(\frac{n(n-1)}{2}\right) = \frac{n(n^2-1)}{3}.$$

Then, the average path length of the path network is

$$L(P_n) = \frac{\sum_{i,j \in V(P_n)} d_{ij}}{n(n-1)} = \frac{\frac{n(n^2-1)}{3}}{n(n-1)} = \frac{n+1}{3}.$$

Then, the agglomeration of the path network is

$$\phi(P_n) = \frac{1}{nL(P_n)} = \frac{1}{n\left(\frac{n+1}{3}\right)} = \frac{3}{n(n+1)}.$$

Thus, the proof holds. ∎

**Theorem 2.1.1.** Let $G = P_n$ $(n > 3)$ of order $n$. Then, for $v \in V(G)$,

$$IMC(v) = \begin{cases} 2/(n+1), & \text{if } v \text{ is an end-node;} \\ 2(2n-1)/(n(n+1)), & \text{otherwise.} \end{cases}$$

**Proof.** There are two cases for computing the importance of the nodes of $P_n$ depending on the types of $P_n$:

Case 1. For an end-node $u$ of $P_n$, the network is agglomerated to a path $P_{n-1}$ of order $n-1$ after the node $u$ is contracted yielding $\phi(P_n'(u)) = \phi(P_{n-1}) = \frac{3}{n(n-1)}$ by Lemma 2.1.1. then, by the use of Lemma 2.1.1, we receive that $IMC(u) = 1 - \frac{\phi(P_n)}{\phi(P_{n-1})} = 1 - \frac{\frac{3}{n(n+1)}}{\frac{3}{n(n-1)}} = \frac{2}{n+1}$.

Case 2. For a node $v$ different than an end-node of $P_n$, the network is agglomerated to a path $P_{n-2}$ of order $n-2$ after the inner node $v$ is contracted yielding

$\phi(P_n'(v)) = \phi(P_{n-2}) = \dfrac{3}{(n-2)(n-1)}$ by Lemma 2.1.1. Then, by the use of Lemma 2.1.1, we

receive that $IMC(v) = 1 - \dfrac{\phi(P_n)}{\phi(P_{n-2})} = 1 - \dfrac{\dfrac{3}{n(n+1)}}{\dfrac{3}{(n-2)(n-1)}} = \dfrac{2(2n-1)}{n(n+1)}$.

The theorem is thus proved. ∎

**Remark 2.1.1.** If a node of a path is contracted, then there are almost no changes- that is the network is agglomerated to a path. Let $u$ be an end-node of degree 1 and $v$ be a node of degree 2 in $P_n$ $(n > 3)$. Being $\deg(v) > \deg(u)$ and $IMC(v) > IMC(u)$, node $v$ has the maximum node degree, the number of nodes after contraction of the node $v$ is the fewest and so the network agglomeration is the highest. Therefore, node $v$ is the most influential node which contributes to network connectivity.

*2.2 Comet networks*

The *comet* $C_{s,t}$ where $s$ and $t$ are positive integers, denotes the tree obtained by identifying the center of the star $K_{1,s}$ with an end-node of $P_t$, the path of order $t$. So $C_{s,1} \cong K_{1,s}$ and $C_{1,t} \cong P_{t+1}$ [4]. Let the center of the star $K_{1,s}$ - that is one end-node of $P_t$ be the node $c$. Label the nodes of $P_t$ sequentially as $v_1, v_2, \ldots, v_{t-1}, c$ where $v_1$ is an end-node and label the nodes of $K_{1,s}$ as $c, u_1, u_2, \ldots, u_s$ where a node $u_i$ $(1 \le i \le s)$ is an end-node.

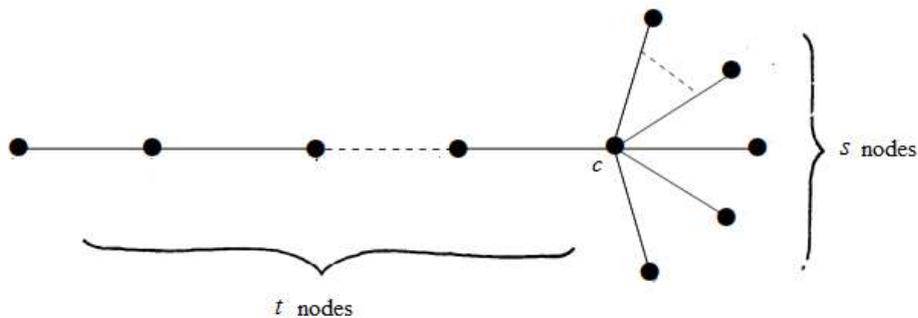

Figure 2.2.1 Comet graph $C_{s,t}$

**Lemma 2.2.1.** *Let $C_{s,t}$ be a comet network of order $s+t$. Then, the agglomeration of the comet network is $\phi(C_{s,t}) = 3(s+t-1)/(t(t+1)(t+3s-1)+6s(s-1))$.*

**Proof.** There are four cases for computing the agglomeration of $C_{s,t}$ depending on the types of the nodes of $C_{s,t}$:

Case 1. For the nodes $v_1, \ldots, v_{t-1}, c$ of $P_t$ in $C_{s,t}$, by Lemma 2.1.1 we have

$$\sum_{\forall u,w \in V(P_t), u \neq w} d_{uw} = \frac{t(t^2-1)}{3}.$$

Case 2. For the nodes $v_1, \ldots, v_{t-1}, c$ of $P_t$ and $u_1, \ldots, u_s$ of $K_{1,s}$ in $C_{s,t}$,

$$\sum_{\forall z \in V(P_t), \forall u_i \in V(K_{1,s})} d_{zu_i} = \sum_{\forall z \in V(P_t)} \sum_{i=1}^{s} d_{zu_i} = s(t) + s(t-1) + \ldots + s(1) = s(1+\ldots+t) = s\left(\frac{t(t+1)}{2}\right).$$

Case 3. For the nodes $u_1, \ldots, u_s$ of $K_{1,s}$ and $v_1, \ldots, v_{t-1}, c$ of $P_t$ in $C_{s,t}$, similar to the Case 2, we receive that

$$\sum_{\forall u_i \in V(K_{1,s}), \forall z \in V(P_t)} d_{u_i z} = \sum_{i=1}^{s} \sum_{\forall z \in V(P_t)} d_{zu_i} = (t+(t-1)+\ldots+(1))(s) = \left(\frac{t(t+1)}{2}\right)(s).$$

Case 4. For the nodes $u_1, \ldots, u_s$ of $K_{1,s}$ in $C_{s,t}$,

$$\sum_{\forall u_i, u_j \in V(K_{1,s}), u_i \neq u_j} d_{u_i u_j} = \sum_{i=1}^{s} \sum_{j=1, j\neq i}^{s} d_{u_i u_j} = s((2)(s-1)).$$

By Cases 1-4, the average path length of the comet network is

$$L(C_{s,t}) = \frac{\sum_{i,j \in V(C_{s,t})} d_{ij}}{(s+t)(s+t-1)} = \frac{\frac{t(t^2-1)}{3} + 2\left(s\left(\frac{t(t+1)}{2}\right)\right) + s(2(s-1))}{(s+t)(s+t-1)}$$

$$L(C_{s,t}) = \frac{t(t^2-1) + 3st(t+1) + 6s(s-1)}{3(s+t)(s+t-1)}.$$

Then, the agglomeration of the comet network is

$$\phi(C_{s,t}) = \frac{1}{(s+t)L(C_{s,t})} = \frac{1}{(s+t)\left(\frac{t(t^2-1)+3st(t+1)+6s(s-1)}{3(s+t)(s+t-1)}\right)}$$

$$\phi(C_{s,t}) = \frac{3(s+t-1)}{t(t+1)(t-1+3s)+6s(s-1)}.$$

Thus, the proof holds. ∎

**Theorem 2.2.1.** Let $G = C_{s,t}$ $(s>2, t>3)$ of order $s+t$. Then, for $v \in V(G)$,

$$IMC(v) = \begin{cases} \dfrac{2t(3s^2-6s+3st-3t+t^2+2)-6s(s-1)}{(t(t+1)(t+3s-1)+6s(s-1))(s+t-2)}, & \text{if } v = v_1; \\[2mm] \dfrac{12st(s+t-3)+2t(2t-5)(t-2)-6(3s-1)(s-1)}{(s+t-3)(t(t+1)(t+3s-1)+6s(s-1))}, & \text{if } v = v_i \ (2 \le i \le t-1); \\[2mm] \dfrac{2(t(s(t+2)+t-1)+3s(s-1))}{t(t+1)(t+3s-1)+6s(s-1)}, & \text{if } v = c; \\[2mm] \dfrac{2t(t^2+6s-7)+6(s-1)(s-2)}{(s+t-2)(t(t+1)(t+3s-1)+6s(s-1))}, & \text{if } v = u_j \ (1 \le j \le s). \end{cases}$$

**Proof.** There are four cases for computing the importance of the nodes of $C_{s,t}$ depending on the types of the nodes of $C_{s,t}$:

Case 1. For the end-node $v_1$ of $P_t$ in $C_{s,t}$, the network is agglomerated to a comet $C_{s,t-1}$ of order $s+t-1$ after the end-node $v_1$ is contracted yielding

$$\phi(C_{s,t}'(v_1)) = \phi(C_{s,t-1}) = \frac{3(s+t-1-1)}{(t-1)(t-1+1)(t-1+3s-1)+6s(s-1)} = \frac{3(s+t-2)}{t(t-1)(t+3s-2)+6s(s-1)}$$

$\phi(C_{s,t}'(v_1)) = \dfrac{3(s+t-2)}{t(t-1)(t+3s-2)+6s(s-1)}$ by Lemma 2.2.1. Then, we receive by the use of

Lemma 2.2.1,

$$IMC(v_1) = 1 - \frac{\phi(C_{s,t})}{\phi(C_{s,t-1})} = 1 - \frac{\dfrac{3(s+t-1)}{t(t+1)(t+3s-1)+6s(s-1)}}{\dfrac{3(s+t-2)}{t(t-1)(t+3s-2)+6s(s-1)}}$$

$$IMC(v_1) = \frac{2t(3s^2-6s+3st-3t+t^2+2)-6s(s-1)}{(t(t+1)(t+3s-1)+6s(s-1))(s+t-2)}.$$

Case 2. For the inner node $v_i$ $(2 \le i \le t-1)$ of $P_t$ in $C_{s,t}$, the network is agglomerated to a comet $C_{s,t-2}$ of order $s+t-2$ after the node $v_i$ is contracted yielding

$$\phi\left(C_{s,t}'(v_i)\right)=\phi(C_{s,t-2})=\frac{3(s+t-2-1)}{(t-2)(t-2+1)(t-2+3s-1)+6s(s-1)}$$

$$\phi\left(C_{s,t}'(v_i)\right)=\frac{3(s+t-3)}{(t-1)(t-2)(t+3s-3)+6s(s-1)}$$

by Lemma 2.2.1. Then, by Lemma 2.2.1, we have

$$IMC(v_i)=1-\frac{\phi(C_{s,t})}{\phi(C_{s,t-2})}=1-\frac{\dfrac{3(s+t-1)}{t(t+1)(t+3s-1)+6s(s-1)}}{\dfrac{3(s+t-3)}{(t-1)(t-2)(t+3s-3)+6s(s-1)}}$$

$$IMC(v_i)=\frac{12st(s+t-3)+2t(2t-5)(t-2)-6(3s-1)(s-1)}{(s+t-3)(t(t+1)(t+3s-1)+6s(s-1))}.$$

Case 3. For the node $c$ of $C_{s,t}$, the network is agglomerated to a path $P_{t-1}$ of order $t-1$ after the node $c$ is contracted yielding $\phi\left(C_{s,t}'(c)\right)=\phi(P_{t-1})=\dfrac{3}{(t-1)(t-1+1)}=\dfrac{3}{t(t-1)}$ by Lemma 2.1.1. Then, we receive by Lemma 2.2.1,

$$IMC(c)=1-\frac{\phi(C_{s,t})}{\phi(P_{t-1})}=1-\frac{\dfrac{3(s+t-1)}{t(t+1)(t+3s-1)+6s(s-1)}}{\dfrac{3}{t(t-1)}}=\frac{2(t(s(t+2)+t-1)+3s(s-1))}{t(t+1)(t+3s-1)+6s(s-1)}.$$

Case 4. For an end-node $u_j$ $(1\le j\le s)$ of $K_{1,s}$ in $C_{s,t}$, the network is agglomerated to a comet $C_{s-1,t}$ of order $s+t-1$ after the end-node $u_j$ is contracted yielding

$$\phi\left(C_{s,t}'(u_j)\right)=\phi(C_{s-1,t})=\frac{3(s-1+t-1)}{t(t+1)(t+3(s-1)-1)+6(s-1)(s-1-1)}$$

$$\phi\left(C_{s,t}'(u_j)\right)=\frac{3(s+t-2)}{t(t+1)(t+3s-4)+6(s-1)(s-2)}$$

by Lemma 2.2.1. Then, by Lemma 2.2.1 we get

$$IMC(u_j)=1-\frac{\phi(C_{s,t})}{\phi(C_{s-1,t})}=1-\frac{\dfrac{3(s+t-1)}{t(t+1)(t+3s-1)+6s(s-1)}}{\dfrac{3(s+t-2)}{t(t+1)(t+3s-4)+6(s-1)(s-2)}}$$

$$IMC(u_j)=\frac{2t(t^2+6s-7)+6(s-1)(s-2)}{(s+t-2)(t(t+1)(t+3s-1)+6s(s-1))}.$$

Thus, the proof of the theorem holds. ∎

**Remark 2.2.1.** The importance of a node in a comet network depends on two factors: one of is its degree and the other is its position. Being $\deg(c) > \deg(v_i) > \deg(v_1) = \deg(u_j)$ yields $IMC(c) > IMC(v_i) > IMC(v_1) > IMC(u_j)$. The degree of the node $c$ is the highest, the number of nodes after contraction is the fewest, and then the network agglomeration is the highest. The node $c$ is the most influential node which contributes to network connectivity. The node $c$ is located in a key position in the network and most of the shortest paths of node pairs pass through the node $c$. If the node $v_i$, or the node $v_1$, or the node $u_j$ in $C_{s,t}$ is contracted, there are almost no changes however being $\deg(v_i) > \deg(v_1) = \deg(u_j)$ yields $IMC(v_i) > IMC(v_1)$ and $IMC(v_i) > IMC(u_j)$ meaning that the higher the degree is and the more critical the position of the node in comet network is, the greater the agglomeration is, and therefore we conclude that node $v_i$ is the more vital node in the network flow. Since the influence of a node in $C_{s,t}$ depends on also its position, even though $\deg(v_1) = \deg(u_j)$, we have that $IMC(v_1) > IMC(u_j)$ meaning that the position of the node $v_1$ is more critical than the position of the node $u_j$ in $C_{s,t}$.

*2.3. Double comet networks*

For $a, b \geq 1$, $n \geq a + b + 2$ by $DC(n, a, b)$ we denote a *double comet*, which is a tree composed of a path containing $n - a - b$ nodes with $a$ pendent nodes attached to one of the ends of the path and $b$ pendent nodes attached to the other end of the path. Thus, $DC(n, a, b)$ has $n$ nodes and $a + b$ leaves [5]. Let $v_1, \ldots, v_a, u_1, \ldots, u_b, w_1, w_2, \ldots, w_{n-a-b}$ be the node set of the double comet $DC(n, a, b)$, which is obtained from a path $P_{n-a-b}$ of nodes $w_1, w_2, \ldots, w_{n-a-b}$ by attaching the pendent nodes $v_1, \ldots, v_a$ to the one end-node $w_1$ of $P_{n-a-b}$ and attaching the pendent nodes $u_1, \ldots, u_b$ to the other end-node $w_{n-a-b}$ of $P_{n-a-b}$.

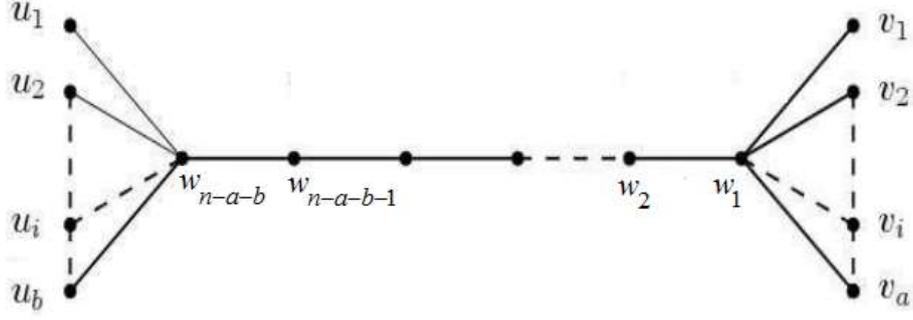

Figure 2.3.1 Double comet graph $DC(n,a,b)$

**Lemma 2.3.1.** *Let $DC(n,a,b)$ be a double comet network of order $n$ $(a,b \geq 1, n-a-b \geq 2)$. Then, the agglomeration of the double comet network is*

$$\phi(DC(n,a,b)) = \frac{3(n-1)}{(n-a-b+1)((n-a-b)(n-a+2b-1)+3a(n-a+b))+6(b(b-1)+a(a-1))}.$$

**Proof.** There are four cases for computing the agglomeration of $DC(n,a,b)$ depending on the types of the nodes of $DC(n,a,b)$:

Case 1. For the nodes $w_1, w_2, \ldots, w_{n-a-b}, u_1, u_2, \ldots, u_b$, let $S = \{w_1, w_2, \ldots, w_{n-a-b}, u_1, u_2, \ldots, u_b\}$. Then, $DC(n,a,b)[S] = C_{b,n-a-b}$ where $C_{b,n-a-b}$ is a comet network of order $n-a$. By Cases 1-4 of Lemma 2.2.1, we have that

$$\sum_{i,j \in V(C_{b,n-a-b})} d_{ij} = \frac{(n-a-b)((n-a-b)^2-1)}{3} + b(n-a-b)(n-a-b+1) + 2b(b-1).$$

Case 2. For the nodes $v_1, v_2, \ldots, v_a$ which are attached to the end-node $w_1$ of $P_{n-a-b}$ in $DC(n,a,b)$,

$$\sum_{i=1}^{a} \sum_{\forall x \in V(C_{b,n-a-b})} d_{v_i x} = \sum_{i=1}^{a} (d_{v_i w_1} + d_{v_i w_2} + \ldots + d_{v_i w_{n-a-b}} + d_{v_i u_1} + d_{v_i u_2} + \ldots + d_{v_i u_b})$$

$$= \sum_{i=1}^{a} (1 + 2 + \ldots + (n-a-b) + b(n-a-b+1)) = \sum_{i=1}^{a} \left( \frac{(n-a-b)(n-a-b+1)}{2} + b(n-a-b+1) \right)$$

$$= a\left( \frac{(n-a-b)(n-a-b+1) + 2b(n-a-b+1)}{2} \right) = \frac{a(n-a-b+1)(n-a+b)}{2}.$$

Case 3. For the nodes $w_1, w_2, \ldots, w_{n-a-b}, u_1, u_2, \ldots, u_b$ of $DC(n,a,b)$, similar to the Case 2, we receive that

$$\sum_{\forall x \in V(C_{b,n-a-b})} \sum_{i=1}^{a} d_{xv_i} = \left(d_{w_1v_1} + d_{w_1v_2} + \ldots + d_{w_1v_a}\right) + \left(d_{w_2v_1} + d_{w_2v_2} + \ldots d_{w_2v_a}\right) + \ldots +$$

$$\left(d_{w_{n-a-b}v_1} + d_{w_{n-a-b}v_2} + \ldots + d_{w_{n-a-b}v_a}\right) + \left(d_{u_1v_1} + d_{u_1v_2} + \ldots + d_{u_1v_a}\right) + \left(d_{u_2v_1} + d_{u_2v_2} + \ldots + d_{u_2v_a}\right) + \ldots +$$

$$\left(d_{u_bv_1} + d_{u_bv_2} + \ldots + d_{u_bv_a}\right)$$

$$= (a)(1) + (a)(2) + \ldots + (a)(n-a-b) + (a)(n-a-b+1) + (a)(n-a-b+1) + \ldots +$$

$$(a)(n-a-b+1)$$

$$= (a)(1 + 2 + \ldots + (n-a-b) + b(n-a-b+1))$$

$$= (a)\left(\frac{(n-a-b)(n-a-b+1)}{2} + b(n-a-b+1)\right)$$

$$= \frac{a(n-a-b+1)(n-a+b)}{2}.$$

Case 4. For the nodes $v_1, v_2, \ldots, v_a$ which are attached to the end-node $w_1$ of $P_{n-a-b}$ in $DC(n,a,b)$,

$$\sum_{i=1}^{a} \sum_{j=1, j \neq i}^{a} d_{v_iv_j} = a\left((a-1)(2)\right) = 2a(a-1).$$

By Cases 1-4, the average path length of the double comet network is

$$L(DC(n,a,b)) = \frac{\sum_{i,j \in V(DC(n,a,b))} d_{ij}}{n(n-1)}$$

$$= \frac{\frac{(n-a-b)\left((n-a-b)^2 - 1\right)}{3} + b(n-a-b)(n-a-b+1) + 2b(b-1) + 2\left(\frac{a(n-a-b+1)(n-a+b)}{2}\right) + 2a(a-1)}{n(n-1)}$$

$$= \frac{(n-a-b+1)\left((n-a-b)(n-a+2b-1) + 3a(n-a+b)\right) + 6(b(b-1) + a(a-1))}{3n(n-1)}.$$

Then, the agglomeration of the double comet network is

$$\phi(DC(n,a,b)) = \frac{1}{nL(DC(n,a,b))}$$

$$= \frac{1}{n\left(\frac{(n-a-b+1)\left((n-a-b)(n-a+2b-1) + 3a(n-a+b)\right) + 6(b(b-1) + a(a-1))}{3n(n-1)}\right)}$$

$$= \frac{3(n-1)}{(n-a-b+1)((n-a-b)(n-a+2b-1)+3a(n-a+b))+6(b(b-1)+a(a-1))}.$$

Thus, the proof holds. ∎

**Theorem 2.3.1.** *Let* $G = DC(n,a,b)$ $(a,b \geq 2, n-a-b \geq 4)$ *of order* $n$, *and let* $n-a-b=k, n-a+2b-1=r, n-a+b=p$. *Then, for* $v \in V(G)$,

$$IMC(v) = \begin{cases} \dfrac{(-k-1)(kr-3p(n-a-1))-6(b(b-1)-(a-1)(2n-a-2))}{(n-2)((k+1)(kr+3ap)+6(b(b-1)+a(a-1)))}, & \text{if } v = v_i \, (1 \leq i \leq a); \\[6pt] \dfrac{(k+1)(k(2n+a-2b-2)+3a(a-b+n-2))+6((b-1)(-b+2n-2)-a(a-1))}{(n-2)((k+1)(kr+3ap)+6(b(b-1)+a(a-1)))}, & \text{if } v = u_j \, (1 \leq j \leq b); \\[6pt] \dfrac{(n-a-2)(k+1)(kr+3ap)+6((n-a-2)a(a-1)-(a+1)b(b-1))-(n-1)(k-1)k(r-1)}{(n-a-2)((k+1)(kr+3ap)+6(b(b-1)+a(a-1)))}, & \text{if } v = w_1; \\[6pt] \dfrac{(n-b-2)((k+1)(kr+3ap)+6b(b-1))-(n-1)(k-1)k(n+2a-b-2)-(b+1)6a(a-1)}{(n-b-2)((k+1)(kr+3ap)+6(b(b-1)+a(a-1)))}, & \text{if } v = w_{n-a-b}; \\[6pt] \dfrac{(n-3)(k+1)(kr+3ap)-(n-1)(k-1)((k-2)(r-2)+3a(p-2))-12(b(b-1)+a(a-1))}{(n-3)((k+1)(kr+3ap)+6(b(b-1)+a(a-1)))}, & \text{if } v = w_t \, (2 \leq t \leq n-a-b-1). \end{cases}$$

**Proof.** There are five cases for computing the importance of the nodes of $DC(n,a,b)$ depending on the types of the nodes of $DC(n,a,b)$:

Case 1. For the pendent node $v_i$ $(1 \leq i \leq a)$ of $DC(n,a,b)$, the network is agglomerated to a double comet $DC(n-1,a-1,b)$ of order $n-1$ after the pendent node $v_i$ is contracted yielding

$$\phi\big(DC(n,a,b)'(v_i)\big) = \phi(DC(n-1,a-1,b))$$

$$= \frac{3(n-2)}{(n-a-b+1)((n-a-b)(n-a+2b-1)+3(a-1)(n-a+b))+6(b(b-1)+(a-1)(a-2))}$$

by Lemma 2.3.1. Then, we receive by the use of Lemma 2.3.1,

$$IMC(v_i) = 1 - \frac{\phi(DC(n,a,b))}{\phi(DC(n-1,a-1,b))}$$

$$= \frac{(-n+a+b-1)((n-a-b)(n-a+2b-1)-3(n-a+b)(n-a-1))-6(b(b-1)-(a-1)(2n-a-2))}{(n-2)((n-a-b+1)((n-a-b)(n-a+2b-1)+3a(n-a+b))+6(b(b-1)+a(a-1)))}.$$

Case 2. For the pendent node $u_j$ $(1 \leq j \leq b)$ of $DC(n,a,b)$, the network is agglomerated to a

double comet $DC(n-1,a,b-1)$ of order $n-1$ after the node $u_j$ is contracted yielding

$$\phi\big(DC(n,a,b)'(u_j)\big) = \phi(DC(n-1,a,b-1))$$

$$= \frac{3(n-2)}{(n-a-b+1)((n-a-b)(n-a+2b-4)+3a(n-a+b-2))+6((b-1)(b-2)+a(a-1))} \text{ by}$$

Lemma 2.3.1. Then, we receive by the use of Lemma 2.3.1,

$$IMC(u_j) = 1 - \frac{\phi(DC(n,a,b))}{\phi(DC(n-1,a,b-1))}$$

$$= 1 - \frac{\dfrac{3(n-1)}{(n-a-b+1)((n-a-b)(n-a+2b-1)+3a(n-a+b))+6(b(b-1)+a(a-1))}}{\dfrac{3(n-2)}{(n-a-b+1)((n-a-b)(n-a+2b-4)+3a(n-a+b-2))+6((b-1)(b-2)+a(a-1))}}$$

$$= \frac{(n-a-b+1)((n-a-b)(2n+a-2b-2)+3a(a-b+n-2))+6((b-1)(-b+2n-2)-a(a-1))}{(n-2)((n-a-b+1)((n-a-b)(n-a+2b-1)+3a(n-a+b))+6(b(b-1)+a(a-1)))}.$$

Case 3. For the node $w_1 \in V(DC(n,a,b))$, the network is agglomerated to a comet $C_{b,n-a-b-1}$ of order $n-a-1$ after the node $w_1$ is contracted yielding

$$\phi\big(DC(n,a,b)'(w_1)\big) = \phi(C_{b,n-a-b-1})$$

$$= \frac{3(b+n-a-b-1-1)}{(n-a-b-1)(n-a-b-1+1)(n-a-b-1+3b-1)+6b(b-1)}$$

$$= \frac{3(n-a-2)}{(n-a-b-1)(n-a-b)(n-a+2b-2)+6b(b-1)} \text{ by Lemma 2.2.1. Then, we receive by}$$

the use of Lemma 2.2.1 and Lemma 2.3.1,

$$IMC(w_1) = 1 - \frac{\phi(DC(n,a,b))}{\phi\big(DC(n,a,b)'(w_1)\big)} = 1 - \frac{\phi(DC(n,a,b))}{\phi(C_{b,n-a-b-1})}$$

$$= 1 - \frac{\dfrac{3(n-1)}{(n-a-b+1)((n-a-b)(n-a+2b-1)+3a(n-a+b))+6(b(b-1)+a(a-1))}}{\dfrac{3(n-a-2)}{(n-a-b-1)(n-a-b)(n-a+2b-2)+6b(b-1)}}$$

$$=\frac{(n-a-2)(n-a-b+1)((n-a-b)(n-a+2b-1)+3a(n-a+b))+6((n-a-2)a(a-1)-(a+1)b(b-1))-(n-1)(n-a-b-1)(n-a-b)(n-a+2b-2)}{(n-a-2)((n-a-b+1)((n-a-b)(n-a+2b-1)+3a(n-a+b))+6(b(b-1)+a(a-1)))}.$$

Case 4. For the node $w_{n-a-b} \in V(DC(n,a,b))$, the network is agglomerated to a comet $C_{a,n-a-b-1}$ of order $n-b-1$ after the node $w_{n-a-b}$ is contracted yielding

$$\phi\left(DC(n,a,b)'(w_{n-a-b})\right)=\phi(C_{a,n-a-b-1})$$

$$=\frac{3(a+n-a-b-1-1)}{(n-a-b-1)(n-a-b-1+1)(n-a-b-1+3a-1)+6a(a-1)}$$

$$=\frac{3(n-b-2)}{(n-a-b-1)(n-a-b)(n+2a-b-2)+6a(a-1)} \quad \text{by Lemma 2.2.1. Then, we receive by}$$

the use of Lemma 2.2.1 and Lemma 2.3.1,

$$IMC(w_{n-a-b})=1-\frac{\phi(DC(n,a,b))}{\phi\left(DC(n,a,b)'(w_{n-a-b})\right)}=1-\frac{\phi(DC(n,a,b))}{\phi(C_{a,n-a-b-1})}$$

$$=1-\frac{\dfrac{3(n-1)}{(n-a-b+1)((n-a-b)(n-a+2b-1)+3a(n-a+b))+6(b(b-1)+a(a-1))}}{\dfrac{3(n-b-2)}{(n-a-b-1)(n-a-b)(n+2a-b-2)+6a(a-1)}}$$

$$=\frac{(n-b-2)((n-a-b+1)((n-a-b)(n-a+2b-1)+3a(n-a+b))+6b(b-1))-(n-1)(n-a-b-1)(n-a-b)(n+2a-b-2)-(b+1)6a(a-1)}{(n-b-2)((n-a-b+1)((n-a-b)(n-a+2b-1)+3a(n-a+b))+6(b(b-1)+a(a-1)))}.$$

Case 5. For a node $w_t$ $(2 \leq t \leq n-a-b-1)$ of $DC(n,a,b)$, the network is agglomerated to a double comet $DC(n-2,a,b)$ of order $n-2$ after the node $w_t$ is contracted yielding

$$\phi\left(DC(n,a,b)'(w_t)\right)=\phi(DC(n-2,a,b))$$

$$=\frac{3(n-2-1)}{(n-2-a-b+1)((n-2-a-b)(n-2-a+2b-1)+3a(n-2-a+b))+6(b(b-1)+a(a-1))}$$

$$=\frac{3(n-3)}{(n-a-b-1)((n-a-b-2)(n-a+2b-3)+3a(n-a+b-2))+6(b(b-1)+a(a-1))} \quad \text{by}$$

Lemma 2.3.1. Then, we receive by the use of Lemma 2.3.1,

$$IMC(w_t)=1-\frac{\phi(DC(n,a,b))}{\phi\left(DC(n,a,b)'(w_t)\right)}=1-\frac{\phi(DC(n,a,b))}{\phi(DC(n-2,a,b))}$$

$$= 1 - \frac{\dfrac{3(n-1)}{(n-a-b+1)((n-a-b)(n-a+2b-1)+3a(n-a+2b))+6(b(b-1)+a(a-1))}}{\dfrac{3(n-3)}{(n-a-b-1)((n-a-b-2)(n-a+2b-3)+3a(n-a+b-2))+6(b(b-1)+a(a-1))}}$$

$$= \frac{(n-3)(n-a-b+1)((n-a-b)(n-a+2b-1)+3a(n-a+b))-(n-1)(n-a-b-1)((n-a-b-2)(n-a+2b-3)+3a(n-a+b-2))-12(b(b-1)+a(a-1))}{(n-3)((n-a-b+1)((n-a-b)(n-a+2b-1)+3a(n-a+b))+6(b(b-1)+a(a-1)))}$$

The theorem is thus proved. ∎

**Remark 2.3.1.** In a double comet network, the network will be more agglomerate- to a comet network after the node $w_{n-a-b}$ or $w_1$ is contracted. If $a > b$, then being $\deg(w_1) > \deg(w_{n-a-b}) > \deg(w_t) > \deg(v_i) = \deg(u_j)$ $(2 \le t \le n-a-b-1, 1 \le i \le a, 1 \le j \le b)$, the number of nodes after contraction of the node $w_1$ is the fewest, and then the network agglomeration is the highest. Being $IMC(w_1) > IMC(w_{n-a-b}) > IMC(w_t) > IMC(u_j) > IMC(v_i)$ yields the node $w_1$ is the most influential node which contributes to network connectivity. The node $w_1$ is located in a vital position in the network and most of the shortest paths of node pairs pass through the node $w_1$. In a similar manner, if $b > a$, then being $\deg(w_{n-a-b}) > \deg(w_1) > \deg(w_t) > \deg(v_i) = \deg(u_j)$, the number of nodes after contraction of the node $w_{n-a-b}$ is the fewest, and then the network agglomeration is the highest. Being $IMC(w_{n-a-b}) > IMC(w_1) > IMC(w_t) > IMC(v_i) > IMC(u_j)$, the node $w_{n-a-b}$ is located in a pivotal position in efficient network flow. If $a = b$, then being $\deg(w_1) = \deg(w_{n-a-b}) > \deg(w_t) > \deg(v_i) = \deg(u_j)$ yields $IMC(w_1) = IMC(w_{n-a-b}) > IMC(w_t) > IMC(v_i) = IMC(u_j)$. If a node $w_t, v_i$, or $u_j$ in $DC(n,a,b)$ is contracted, then there are almost no changes-the network will agglomerate to a double comet again. However being $\deg(w_t) > \deg(v_i)$ and $\deg(w_t) > \deg(u_j)$ yields $IMC(w_t) > IMC(v_i)$ and $IMC(w_t) > IMC(u_j)$ since the importance of a node depends on also its position in the network. Therefore, the node $w_t$ is more influential than the nodes $v_i$ and $u_j$.

## 2.4. Lollipop networks

The *lollipop network* $L_{n,d}$ is a graph obtained from a complete graph $K_{n-d}$ and a path $P_d$ $(d>1)$, by joining one of the end-nodes of $P_d$ to all the nodes of $K_{n-d}$ [7]. Let the nodes of $P_d$ be $v_1,\ldots,v_d$ where $v_1$ and $v_d$ are the end-nodes of $P_d$ and $\deg_{L_{n,d}}(v_1)=1$, $\deg_{L_{n,d}}(v_d)=n-d+1$, and let the nodes of $K_{n-d}$ be $u_1,\ldots,u_{n-d}$.

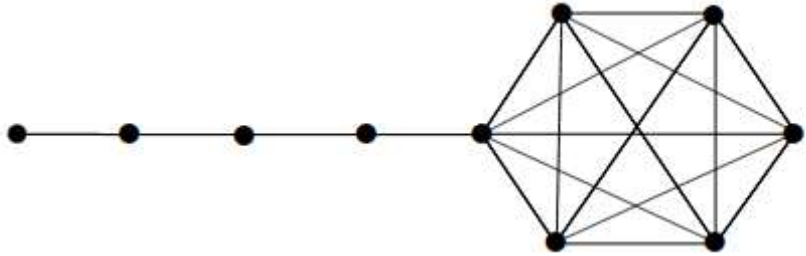

Figure 2.4.1 Lollipop graph $L_{n,d}$ for $n=10$ and $d=5$

**Lemma 2.4.1.** Let $L_{n,d}$ be a lollipop network of order $n$. Then, the agglomeration of the lollipop network is $\phi(L_{n,d}) = 3(n-1)\big/\big(3(n-d)(d^2-1+n)+d(d^2-1)\big)$.

**Proof.** There are four cases for computing the agglomeration of $L_{n,d}$ depending on the types of the nodes of $L_{n,d}$:

Case 1. For the nodes $v_1,\ldots,v_d$ of $P_d$ in $L_{n,d}$, $\displaystyle\sum_{\forall v_i,v_j \in V(P_n), v_i \neq v_j} d_{v_i v_j} = \sum_{i=1}^{d}\sum_{j=1, j\neq i}^{d} d_{v_i v_j}$. By the proof of

Lemma 2.1.1, $\displaystyle\sum_{i=1}^{d}\sum_{j=1, j\neq i}^{d} d_{v_i v_j} = \frac{d(d^2-1)}{3}$.

Case 2. For the nodes $v_1,\ldots,v_d$ of $P_d$ and $u_1,\ldots,u_{n-d}$ of $K_{n-d}$ in $L_{n,d}$,

$$\sum_{\forall v_i \in V(P_n), \forall u_j \in V(K_{n-d})} d_{v_i u_j} = \sum_{i=1}^{d}\sum_{j=1}^{n-d} d_{v_i u_j} = d(n-d)+(d-1)(n-d)+\ldots+(1)(n-d)$$

$$=(n-d)(1+\ldots+d)=(n-d)\left(\frac{d(d+1)}{2}\right).$$

Case 3. For the nodes $u_1,\ldots,u_{n-d}$ of $K_{n-d}$ and $v_1,\ldots,v_d$ of $P_d$ in $L_{n,d}$, similar to the Case 2,

we obtain $\sum_{\forall u_i \in V(K_{n-d}), \forall v_j \in V(P_n)} d_{u_i v_j} = \sum_{i=1}^{n-d}\sum_{j=1}^{d} d_{u_i v_j} = (d+(d-1)+\ldots+(1))(n-d) = \frac{d(d+1)}{2}(n-d)$.

Case 4. For the nodes $u_1,\ldots,u_{n-d}$ of $K_{n-d}$ in $L_{n,d}$,

$$\sum_{\forall u_i, u_j \in V(K_{n-d}), u_i \neq u_j} d_{u_i u_j} = \sum_{i=1}^{n-d}\sum_{j=1, j\neq i}^{n-d} d_{u_i u_j} = (n-d)((n-d-1)(1)).$$

By Cases 1-4, the average path length of the lollipop network is

$$L(L_{n,d}) = \frac{\sum_{i,j \in V(L_{n,d})} d_{ij}}{(n)(n-1)} = \frac{\frac{d(d^2-1)}{3}+2\left((n-d)\left(\frac{d(d+1)}{2}\right)\right)+(n-d)(n-d-1)}{n(n-1)}$$

$$= \frac{d(d^2-1)+3d(n-d)(d+1)+3(n-d)(n-d-1)}{3n(n-1)}.$$

Then, the agglomeration of the lollipop network is

$$\phi(L_{n,d}) = \frac{1}{(n)L(L_{n,d})} = \frac{1}{n\left(\frac{d(d^2-1)+3d(n-d)(d+1)+3(n-d)(n-d-1)}{3n(n-1)}\right)}$$

$$= \frac{3(n-1)}{3(n-d)(d^2-1+n)+d(d^2-1)}.$$

Thus, the proof holds. ∎

**Theorem 2.4.1.** Let $G = L_{n,d}$ $(d>3, n-d>1)$ of order $n$. Then, for $v \in V(G)$,

$$IMC(v) = \begin{cases} \dfrac{3(n-d)((n-1)(2d-1)-d^2)+d(d-1)(3n-d-4)}{(n-2)(3(n-d)(d^2-1+n)+d(d^2-1))}, & \text{if } v = v_1; \\[2ex] \dfrac{6(n-d)(2(d-1)(n-1)-d^2)+2(d-1)(3(d-1)(n-1)-d(d+1))}{(n-3)(3(n-d)(d^2-1+n)+d(d^2-1))}, & \text{if } v = v_i\, (2 \leq i \leq d-1); \\[2ex] \dfrac{3(n-d)(d^2-1+n)+d(d-1)(d-n+2)}{3(n-d)(d^2-1+n)+d(d^2-1)}, & \text{if } v = v_d; \\[2ex] \dfrac{(n-d)(d(2d-1)+3(n-1))}{3(n-d)(d^2-1+n)+d(d^2-1)}, & \text{if } v = u_j\, (1 \leq j \leq n-d). \end{cases}$$

**Proof.** There are four cases for computing the importance of the nodes of $L_{n,d}$ depending on the types of the nodes of $L_{n,d}$:

Case 1. For the node $v_1$ of $P_d$, the network is agglomerated to a lollipop network $L_{n-1,d-1}$ of order $n-1$ after the node $v_1$ is contracted yielding

$$\phi\left(L_{n,d}{}'(v_1)\right) = \phi\left(L_{n-1,d-1}\right) = \frac{3(n-1-1)}{3(n-1-(d-1))\left((d-1)^2-1+n-1\right)+(d-1)\left((d-1)^2-1\right)}$$

$$= \frac{3(n-2)}{3(n-d)\left(d^2-2d+n-1\right)+(d)(d-1)(d-2)}$$

by Lemma 2.4.1. Then, we receive by Lemma 2.4.1,

$$IMC(v_1) = 1 - \frac{\phi(L_{n,d})}{\phi(L_{n-1,d-1})} = 1 - \frac{\dfrac{3(n-1)}{3(n-d)(d^2-1+n)+d(d^2-1)}}{\dfrac{3(n-2)}{3(n-d)(d^2-2d+n-1)+d(d-1)(d-2)}}$$

$$= \frac{3(n-d)\left((n-1)(2d-1)-d^2\right)+d(d-1)(3n-d-4)}{(n-2)\left(3(n-d)(d^2-1+n)+d(d^2-1)\right)}.$$

Case 2. For the node $v_i$ $(2 \leq i \leq d-1)$ of $P_d$, the network is agglomerated to a lollipop network $L_{n-2,d-2}$ of order $n-2$ after the node $v_i$ is contracted yielding

$$\phi\left(L_{n,d}{}'(v_i)\right) = \phi\left(L_{n-2,d-2}\right) = \frac{3(n-2-1)}{3(n-2-(d-2))\left((d-2)^2-1+n-2\right)+(d-2)\left((d-2)^2-1\right)}$$

$$= \frac{3(n-3)}{3(n-d)\left(d^2-4d+1+n\right)+(d-1)(d-2)(d-3)}$$

by Lemma 2.4.1. Then, by Lemma 2.4.1, we have that

$$IMC(v_i) = 1 - \frac{\phi(L_{n,d})}{\phi(L_{n-2,d-2})} = 1 - \frac{\dfrac{3(n-1)}{3(n-d)(d^2-1+n)+d(d^2-1)}}{\dfrac{3(n-3)}{3(n-d)(d^2-4d+1+n)+(d-1)(d-2)(d-3)}}$$

$$= \frac{6(n-d)\left(2(d-1)(n-1)-d^2\right)+2(d-1)\left(3(d-1)(n-1)-d(d+1)\right)}{(n-3)\left(3(n-d)(d^2-1+n)+d(d^2-1)\right)}.$$

Case 3. For the node $v_d$ of $P_d$, the network is agglomerated to a path $P_{d-1}$ of order $d-1$ after the node $v_d$ is contracted yielding $\phi(L_{n,d}'(v_d)) = \phi(P_{d-1}) = \frac{3}{(d-1)(d-1+1)} = \frac{3}{d(d-1)}$ by Lemma 2.1.1. Then, by Lemma 2.4.1, we receive that

$$IMC(v_d) = 1 - \frac{\phi(L_{n,d})}{\phi(P_{d-1})} = 1 - \frac{\frac{3(n-1)}{3(n-d)(d^2-1+n)+d(d^2-1)}}{\frac{3}{d(d-1)}}$$

$$= \frac{3(n-d)(d^2-1+n)+d(d-1)(d-n+2)}{3(n-d)(d^2-1+n)+d(d^2-1)}.$$

Case 4. For the node $u_j$ $(1 \leq j \leq n-d)$ of $K_{n-d}$, the network is agglomerated to a path $P_d$ of order $d$ after the node $u_j$ is contracted yielding $\phi(L_{n,d}'(u_j)) = \phi(P_d) = \frac{3}{d(d+1)}$ by Lemma 2.1.1. Then, by Lemma 2.4.1, we have that

$$IMC(u_j) = 1 - \frac{\phi(L_{n,d})}{\phi(P_d)} = 1 - \frac{\frac{3(n-1)}{3(n-d)(d^2-1+n)+d(d^2-1)}}{\frac{3}{d(d+1)}}$$

$$= \frac{(n-d)(d(2d-1)+3(n-1))}{3(n-d)(d^2-1+n)+d(d^2-1)}.$$

The theorem is thus proved. ∎

**Remark 2.4.1.** The importance of a node in a lollipop network depends on two factors as one of is its degree and the other is its position. In a lollipop network, the network is more agglomerate-to a path after the node $v_d$ is contracted. Being $\Delta(L_{n,d}) = \deg(v_d)$, $\delta(L_{n,d}) = \deg(v_1)$, the degree of the node $v_d$ is the highest, the number of nodes after contraction is the fewest, and then the network agglomeration is the highest-to a path, and $IMC(v_d) > IMC(v_1)$, $IMC(v_d) > IMC(v_i)$, $IMC(v_d) > IMC(u_j)$, the node $v_d$ is the most influential node which contributes to network connectivity. Accordingly, the node $v_d$ is located in a pivotal position in the efficient network flow and most of the shortest paths of node pairs pass through the node $v_d$. If the node $v_i$ or $v_1$ in $L_{n,d}$ is contracted, then there are almost no changes however being $\deg(v_i) > \deg(v_1)$ yields $IMC(v_i) > IMC(v_1)$. If the node $u_j$ in $L_{n,d}$ is contracted, then the network is again more agglomerate-to a path however being $\deg(v_d) > \deg(u_j)$ yields $IMC(v_d) > IMC(u_j)$, but $IMC(u_j) > IMC(v_1)$. Since the influence of a node in $L_{n,d}$ depends on also its position, we can conclude that the positions of the nodes $u_j$ and $v_i$ are more critical than the position of the node $v_1$ in $L_{n,d}$. If $n-d=2$, then even though $\deg(u_j) = \deg(v_i)$, we have that $IMC(v_i) > IMC(u_j)$ meaning that the node $v_i$ is more influential in the network flow than the node $u_j$. If $n-d>2$, then we have that $\deg(u_j) > \deg(v_i)$. The higher the degree is and the more critical the position of the node in the network $L_{n,d}$ is, the greater the agglomeration is, and therefore we conclude that $IMC(u_j) > IMC(v_i)$ except the special cases for which $n=7, d=4$ and $n=8, d=5$. But in general we can say that clearly the node $u_j$ is more vital in the network than the node $v_i$.